\documentclass[12pt]{iopart}
\usepackage{epsf,epsfig}
\begin{document}
\jl{31}
\newcommand{\dd}{{\rm d}}
\title{A Dynamical Study of the Friedmann Equations}
\author{Jean-Philippe Uzan$^\dag$ and Roland Lehoucq$^\ddag$}
\address{\dag Laboratoire de Physique Th\'eorique, CNRS--UMR 8627,\\
          B\^at. 210, Universit\'e Paris XI, F--91405 Orsay (France)}
 \address{\ddag CE-Saclay/DSM/DAPNIA/Service d'Astrophysique,\\
           F--91191 Gif sur Yvette cedex (France).}
\begin{abstract}
Cosmology is an attracting subject for students but usually difficult to
deal with if general relativity is not known. In this article, we first
recall the Newtonian derivation of the Friedmann equations which govern
the dynamics of our universe and discuss the validity of such a
derivation. We then study the equations of evolution of the universe in
terms of a dynamical system. This sums up the different behaviors of
our universe and enables to address some cosmological problems.
\end{abstract}
\pacs{???}

\section{Introduction}\label{I}

In this article, we want to present a pedagogical approach to the
equations governing the evolution of the universe, namely the Friedmann
equations. Indeed, the derivation of this equations is intrinsically
relativistic. Although in Newtonian theory, the universe must be static,
Milne~\cite{milne34} and McCrea and Milne~\cite{mccrea34} showed that,
surprisingly, the Friedmann equations can be derived from the simpler
Newtonian theory. In section~\ref{II}, we recall their derivation
(\S~\ref{II1}) for a universe filled with pressureless matter and then
discuss the introduction of a cosmological constant
(\S~\ref{II2}). Indeed, it is puzzling that the Newtonian theory and
general relativity give the same results; we briefly discuss this issue in
\S~\ref{II3}.

Once we have interpreted the Friedmann equations, we study them as a
dynamical system. The first authors to consider such an approach were
Stabell and Refsdal~\cite{stabell66} who investigated the
Friedmann--Lema\^{\i}tre model with a pressureless fluid. This was
then generalised to a fluid with any equation of
state~\cite{madsen88,madsen92}. Then, this technique was intensively
used to study the isotropisation of homogeneous models (see
e.g. \cite{goliath99} and references therein).  For a general
description of the use of dynamical systems in cosmology, we refer to
the book by Wainwright and Ellis~\cite{bookwe} where most of the
techniques are detailed.  Our purpose here, is to present such an
analysis for a fluid with any equation of state and including a
cosmological constant while staying as pedagogical as possible. In
section \S~\ref{III}, we rewrite the Friedmann equations under a form
easier to handle with and we extract the dynamical system to study. We
then determine the fixed points of this system and discuss their
stability. We illustrate this analytic study by a numerical
integration of this set of equations (\S~\ref{IV}) and finish by a
discussion about the initial conditions explaining the current
observed state of our universe (\S~\ref{V}).

\section{A Newtonian derivation of the Friedmann equation}\label{II}

We follow the approach by Milne~\cite{milne34} and McCrea and
Milne~\cite{mccrea34} and the reader is referred to~\cite{harrison00} for
further details.

\subsection{General derivation}\label{II1}

We consider a sphere of radius $R$ filled with a pressureless fluid
($P=0$) of uniform (mass) density $\rho$ free--falling under its own
gravitational field in an otherwise empty Euclidean space.
We decompose the coordinate ${\bf x}$ of any  particle of the fluid as
\begin{equation}\label{1}
{\bf x}=a(t){\bf r}
\end{equation}
where ${\bf r}$ is a constant vector referred to as the {\it comoving
coordinate}, $t$ is the time coordinate and $a$ the scale factor. We
choose $a$ to have the dimension of a length and $r$ to be
dimensionless. It implies that the sphere undergoes a self similar
expansion or contraction and that no particle can cross another
one. Indeed the edge of the sphere is also moving as
\begin{equation}\label{2}
R(t)=a(t)R_0.
\end{equation}
Assume that while sitting on a particle labelled $i$ we are observing
a particle labelled $j$; we see it drift with the relative velocity 
\begin{equation}\label{3}
{\bf v}_{ij}=\dot a\left({\bf r}_j-{\bf r}_i\right)=H
{\bf x}_{ij}
\end{equation}
where a dot refers to a time derivative, $H\equiv\dot a/a$ and ${\bf
x}_{ij}\equiv\left({\bf r}_j-{\bf r}_i\right)$. As a consequence, any
particle $i$ sees any other particle $j$ with a radial velocity
proportional to its distance and the expansion is isotropic with respect
to any point of the sphere, whatever the function $a(t)$. But, note
that this does not imply that all particles are equivalent (as will be
discussed later).

To determine the equation of motion of any particle of this expanding
sphere, we first write the equation of matter conservation stating that
the mass within any comoving volume is constant (i.e. $\rho x^3 \propto r^3$)
implying that
\begin{equation}\label{4}
\rho(t)\propto a^{-3}(t),
\end{equation}
which can also be written under the form
\begin{equation}\label{5}
\dot\rho + 3H\rho = 0.
\end{equation}
Note that Eq.~(\ref{5}) can also be deduced from the more general
conservation equation $\partial_t\rho + \nabla_x{\bf j}=0$ with ${\bf
j}=\rho{\bf v}$, ${\bf v}=H{\bf x}$ and $\nabla_x{\bf x}=3$. 

To determine the equation of evolution of the scale factor $a$, we
first compute the gravitational potential energy $E_G$ of a particle
of masse $m$ by applying the Gauss law
\begin{equation}\label{6}
E_G=-\frac{G M(<x) m}{x}
\end{equation}
where $G$ is the Newton constant and $M(<x)$ the mass within the sphere
of radius $x$ given by
\begin{equation}\label{7}
M(<x)=\frac{4\pi}{3}\rho x^3.
\end{equation}
We then need to evaluate its kinetic energy $E_K$ which takes the
simple form
\begin{equation}\label{8}
E_K = \frac{1}{2}m\dot x^2.
\end{equation}
The conservation of the total energy $E = E_G + E_K$ implies, after the
use of the decomposition (\ref{1}) and a simplification by $r$, that
\begin{equation}\label{8a}
\left(\frac{\dot a}{a}\right)^2=\frac{8\pi G}{3}\rho - \frac{Kc^2}{a^2}
\end{equation}
where $K$ is a dimensionless constant (which can depend on $r$) given by
$K=-2E/(mc^2r^2)$\footnote{This scaling of $K$ with $r$ is imposed by
the requirement that the expansion is self--similar (Eq.~\ref{1}) and
that no shell of labeled $r$ can cross a shell of label $r'>r$.}. 

\subsection{Introducing a cosmological constant}\label{II2}

In the former derivation, the gravitational potential on any particle
inside the sphere is proportional to the distance $x^2$. Any other
force deriving from a potential proportional to $x^2$ will mimic a
gravitational effect. A force deriving from the potential energy $E_\Lambda$
defined by
\begin{equation}\label{10}
E_\Lambda=-m\frac{\Lambda c^2}{6}x^2
\end{equation}
where $\Lambda$ is a constant was introduced by Einstein in 1917. As
in the previous section, writing that the total energy
$E=E_K+E_G+E_\Lambda$ is constant leads
to the equation of motion
\begin{equation}\label{11}
\left(\frac{\dot a}{a}\right)^2=\frac{8\pi G}{3}\rho-\frac{Kc^2}{a^2}
+\frac{\Lambda c^2}{3}.
\end{equation}
From (\ref{10}), we deduce that $\Lambda$ has the dimension of an
inverse squared length. The total force on a particle is
\begin{equation}\label{12}
{\bf F}=m\left(-\frac{4\pi G}{3}\rho+\frac{\Lambda c^2}{3}\right){\bf x}
\end{equation}
from which it can be concluded that (i) it opposes gravity if
$\Lambda$ is positive and that (ii) it can be tuned so that ${\bf
F}=0$ leading to $\dot a=0$ and $\rho=$constant if
\begin{equation}
\Lambda =\frac{4\pi G}{c^2}\rho.
\end{equation}
This enables to recover a static autogravitating sphere hence leading
to a model for a static universe. The force deriving from $E_\Lambda$ is
analogous to the one exerted by a spring of negative constant.

To finish, we recall on table~\ref{tableunit} the dimension of all the
quantities used in the former sections, mainly to compare with standard
textbooks in which the choice $c=1$ is usually made.

\begin{table}
\caption{Units of the quantities introduced in the article. $M$, $L$
and $T$ stand respectively for mass, length and time units.}
\label{tableunit}
\begin{indented}
\item[]
\begin{tabular}{ccccccccc}
\br
$a$ & $r$ & ${\bf v}$ & $\rho$ & $P$ & $E$ & $H$ & $\Lambda$ & $K$ \\
\mr
$L$ & $-$ & $L.T^{-1}$ & $M.L^{-3}$ & $M.L^{-1}.T^{-2}$ &
$M.L^{2}.T^{-2}$ & $T{-1}$ & $L^{-2}$ & $-$ \\
\br
\end{tabular}
\end{indented}
\end{table}

\subsection{Discussion}\label{II3}

From this Newtonian approach, the equation of evolution of the
universe identified with this gravitating sphere are thus given by
equation (\ref{5}) and (\ref{11}). These are two differential
equations for the two variables $a(t)$ and $\rho(t)$ which can be
solved once the two parameters $K$ and $\Lambda$ have been chosen.

In the context of general relativity, one can deduce the law of
evolution for the scale factor of the universe $a$ which is given by
the Friedmann equations
\begin{eqnarray}
H^2&=&\frac{\kappa}{3}\rho-\frac{Kc^2}{a^2}+\frac{\Lambda c^2}{3}
\label{fried1}\\
\frac{\ddot a}{a}&=&-\frac{\kappa}{6}(\rho+3\frac{P}{c^2})+\frac{\Lambda
c^2}{3}
\label{fried2}
\end{eqnarray}
with $\kappa\equiv8\pi G$ and the conservation equation
\begin{equation}
\dot\rho+3H(\rho+\frac{P}{c^2})=0.\label{eq_cons}
\end{equation}
Eq.~(\ref{fried1}) reduces to (\ref{11}) and, Eq.~(\ref{eq_cons}) to
(\ref{5}) when $P=0$. The equation (\ref{fried2}) is redundant and can
be deduced from the two others. Note that now Eq.~(\ref{eq_cons}) is
also a conservation equation but with the mass flux ${\bf
j}=(\rho+P/c^2){\bf v}$. This can be interpreted by remembering that the
first law of thermodynamics for an adiabatic system~\cite{tolman} takes
the form
\begin{equation}\label{thermo1}
\dot E +P\dot V=0
\end{equation}
where $E=\rho V c^2$ is the energy contained in the physical volume $V$
(scaling as $a^{3}$).

The first thing to stress is that equations (\ref{5}) and (\ref{11}) do
not depend on the radius $R_0$ of the sphere. It thus seems that we can
let it go to infinity without changing the conclusions and hence
explaining why we recover the Friedmann equations.  This was the
viewpoint adopted by Milne~\cite{milne34} and McCrea and
Milne~\cite{mccrea34}. This approach leads to some problems. First, it
has to be checked that the Gauss theorem still applies after taking the
limit toward infinity (i.e. one has to check that the integrals and the
limit commute). This imposes that $\rho$ decreases fast enough with $r$
and thus that there is indeed a center. Equivalently, as pointed out by
Layzer~\cite{layzer54}, the force on any particle of an infinite
homogeneous distribution is undetermined (the integral over the angles
is zero while the integral over the radial coordinate is infinite). The
convergence of the force requires either the mass distribution to be
finite (in which case it can be homogeneous) or to be inhomogeneous if
it is infinite. The issue of the finiteness of the universe has been
widely discussed and a clear presentation of the evolution of ideas in
that respect are presented in~\cite{ha2}.  Second, for distances of
cosmological interests, i.e. of some hundred of Megaparsec, the
recession speed of the particles of the sphere are of order of some
fraction of the speed of light. One will thus require a (special)
relativistic treatment of the expanding sphere.  Third, the
gravitational potential grows with the square of the radius of the
sphere but it can not become too large otherwise, due to the virial
theorem, the velocities would exceed the speed of light.

It was then proposed~\cite{callan65} that such an expanding sphere may
describe a region the size of which is small compared with the size of
the observable universe (i.e. of the Hubble size). Since all regions
of a uniform and isotropic universe expand the same way, the study of
a small region gives information about the whole universe (but this
does not solve the problem of the computation of the gravitational
force).

The center seems to be a privileged points since it is the
only point to be at rest with respect to the absolute frame. But, one can
show that the spacetime background of Newtonian mechanics is invariant
under a larger group than the traditionally described Galilean group. As
shown by Milne~\cite{milne34}, McCrea and Milne~\cite{mccrea34} and
Bonnor~\cite{bonnor57} (see also Carter and Gaffet~\cite{carter87} for a
modern description) it includes the set of all time-dependent space
translations
$$
x^i\rightarrow x^i+z^i(t)
$$
where $z^i(t)$ are arbitrarily differentiable functions depending only
on the time coordinate $t$. This group of transformation is intermediate
between the Galilean group and the group of all diffeomorphisms under
which the Einstein theory in invariant. Thanks to this invariance group,
each point can be chosen as a center around which there is local
isotropy and homogeneity but the isotropy is broken by the existence of
the boundary of the sphere (i.e. all observer can believe living at the
center as long as he/she does not observe the boundary of the expanding
sphere).

There are also conceptual differences between the Newtonian cosmology
and the relativist cosmology. In the former we have a sphere of particle
moving in a static and absolute Euclidean space and the time of
evolution of the sphere is disconnected from the absolute time $t$. For
instance in a recollapsing sphere, the time will go on flowing even
after the crunch of the sphere. In general relativity, space is
expanding and the particles are comoving. We thus identify an expanding
sphere in a fixed background and an expanding spacetime with fixed
particles. As long as we are dealing with a pressureless fluid, this is
possible since there is no pressure gradient and each point of the
sphere can be identify with one point of space (in fact, with an
absolute time we are working in a synchronous reference frame and we
want it to be also comoving, which is possible only if
$P=0$~\cite{landau}). Moreover, the pressure term in the Friedmann
equations cannot trivially be recovered from the Newtonian argument. As
shown, one gets the correct Friedmann equations if one starts from the
conservation law including pressure (and derived from the first law of
thermodynamics) and the conservation of energy. But if one were starting
from the Newton law relating force (\ref{12}) and acceleration ($m\ddot
a{\bf x}$), the term containing the pressure in (\ref{fried2}) would not
have been recovered; one should have added an extra pressure
contribution ${\bf F}_P=-4\pi GmP{\bf x}/c^2$ which can not be
guessed. This is a consequence that in general relativity any type of
energy has a gravitational effect. In a way it is a ``miracle'' that the
equation (\ref{fried1}) does not depend on $P$, which makes it possible
to derive from the Newtonian conservation of energy. Beside it has also to be
stressed that the Newtonian derivation of the Friedmann equations by
Milne came after Friedmann and Lema\^{\i}tre demonstrated the validity
of the Friedmann equations for an unbounded homogeneous distribution of
matter (using general relativity). It has to be pointed out that these
Newtonian models can not explain all the observational relations since,
contrary to general relativity, they do not incorporate a theory of
light propagation. As outlined by Lazer~\cite{layzer54} one can sometime
legitimately treat a part of the (dust) expanding universe as an
isolated system in which case the Newtonian treatment is correct, which
makes McCrea~\cite{mccrea55} conclude that this is an {\it indication}
that Einstein's law of gravity must admit the same {\it interpretation}
as that of Newton's in the case of a spherically symmetric mass
distribution. Note that the structural similarity of Einstein and Newton
gravity were put forward by Cartan~\cite{cartan} who showed that these
two theories are much closer that one naively thought and, in that
framework (which goes far beyond our purpose) one can work out a correct
derivation of the Friedmann equations (see e.g.~\cite{strau}).

The most important outcome of the Newtonian derivation of the Friedmann
equations is that it allows to interpret equation (\ref{fried1}) in
terms of the conservation of energy; the term in $H^2$ represents the
kinetic energy, the term in $\kappa\rho/3$ the gravitational potential
energy, the term in $\Lambda/3$ the energy associated with the
cosmological constant and the term in $K$ the total energy of the
system. The properties of the spatial sections (i.e. of the three
dimensional spaces of constant time) are related to the sign of $K$ and
can be compared with the property of the trajectories of the point of
the sphere which are related to the sign of the total energy $E$; we sum
up all these properties on table~\ref{tab2}.

\begin{table}
\caption{Comparison of the nature of the Newtonian trajectory and
of the structure of space according to the value of
the constant $K$ in Eq.~(\ref{11}).}
\label{tab2}
\begin{indented}
\item[]
\begin{tabular}{lccc}
\br
$E$        & $>0$       & $0$       & $<0$    \\
\mr
Trajectory & hyperbolic & parabolic & elliptic\\
           & unbounded  & unbounded & bounded \\
\br
$K$        & $<0$       & $0$       & $>0$     \\
\mr
Spatial section & infinite & infinite & finite \\
\br
\end{tabular}
\end{indented}
\end{table}

\section{The Friedmann equations as a dynamical system}\label{III}

The Friedmann equations (\ref{fried1}--\ref{fried2}) and the
conservation equation (\ref{eq_cons}) form a set of two
independent equations for three variables ($a$, $P$ and $\rho$).
The usual approach is to solve this system by specifying the
matter content of the universe mainly by assuming an equation of
state of the form
\begin{equation}
P=(\gamma-1)\rho c^2
\end{equation}
where $\gamma$ may depend on $\rho$ and
thus on time. For a pressureless fluid (modelling for instance a fluid
of galaxies) $\gamma=1$ and for a fluid of radiation (such as photon,
neutrino,...) $\gamma=4/3$. We assume that $\gamma\not=0$ since such a
type of matter is described by the cosmological constant and singled out
from ``ordinary" matter and that $\gamma\not=2/3$ since such a type of
matter mimics the curvature term and is thus incorporated with it.

One can then first integrate (\ref{eq_cons}) rewritten as
$\dd\rho/(\gamma\rho)=-3\dd a/a$ to get the function $\rho(a)$ which,
in the case where $\gamma$ is constant, yields
\begin{equation}\label{Cgamma}
\rho(a)=C a^{-3\gamma}
\end{equation}
where $C$ is a positive constant of integration, and then insert
the solution for $\rho(a)$ in Eq.~(\ref{fried1}) to get a closed
equation for the scale factor $a$ (see e.g.~\cite{peebles93} for such
an approach and \cite{faraoni99} for an alternative and pedagogical
derivation).

In this section, we want to present another approach in which the
Friedmann equations are considered as a dynamical system and to
determine its phase space.

\subsection{Derivation of the system}\label{III1}

The first step is to rewrite the set of dynamical equations with
the three new variables $\Omega$, $\Omega_\Lambda$ and $\Omega_K$
defined as
\begin{eqnarray}
\Omega&\equiv& \frac{\kappa\rho}{3H^2} \, ,\label{O}\\
\Omega_\Lambda&\equiv& \frac{\Lambda c^2}{3H^2}\, , \label{Ol}\\
\Omega_K&\equiv&-\frac{Kc^2}{a^2H^2}\, .\label{Ok}
\end{eqnarray}
They respectively represent the relative amount of energy density
present in the matter distribution, cosmological constant and
curvature. $\Omega$ has to be positive and there is no constraint
on the sign of both $\Omega_\Lambda$ and $\Omega_K$. With these
definitions, it is straightforward to deduce from (\ref{fried1})
that
\begin{equation}
\Omega+\Omega_\Lambda+\Omega_K=1.\label{sum}
\end{equation}
Using that $\dot H=\ddot a/a-H^2$, expressing $\ddot a/a$ from Eq.
(\ref{fried2}) and $H^2$ from Eq. (\ref{fried1}), we deduce that
\begin{equation}
\frac{\dot H}{H^2}=-(1+q)\label{hub2}
\end{equation}
where the deceleration parameter $q$ is defined by
\begin{equation}
q\equiv\frac{3\gamma-2}{2}(1-\Omega_K)-\frac{3\gamma}{2}\Omega_\Lambda.
\label{def_q}
\end{equation}
It is useful to rewrite the full set of equations by introducing the
new dimensionless time variable $\eta\equiv{\rm ln}(a/a_0)$, $a_0$ being
for instance the value of $a$ today. The derivative of
any quantity $X$ with respect to $\eta$, $X'$, is then related to its
derivative with respect to $t$ by $X'=\dot X/H$. The equation of
evolution of the Hubble parameter (\ref{hub2}) takes the form
\begin{equation}
H'=-(1+q)H.\label{hub3}
\end{equation}
Now, differentiating $\Omega$, $\Omega_\Lambda$ and $\Omega_K$ with
respect to $\eta$, using Eq. (\ref{hub3}) to express $H'$, $a'=a$ and
Eq. (\ref{eq_cons}) to express $\rho'=-3\gamma\rho$, we obtain the
system
\begin{eqnarray}
\Omega'&=& (2q+2-3\gamma)\Omega\label{Oprim}\\ \Omega_\Lambda'&=&
2(1+q)\Omega_\Lambda \label{Olprim}\\
\Omega_K'&=&2q\Omega_K\label{Okprim}
\end{eqnarray}
and it is trivial to check that
$\Omega'+\Omega_\Lambda'+\Omega_K'=0$ as expected form
(\ref{sum}).

Indeed, it is useless to study the full system
(\ref{hub3}--\ref{Okprim}) (i) since $H$ does not enter the set of
equations (\ref{Oprim}--\ref{Okprim}) and is solely determined by
Eq. (\ref{hub3}) once this system has been solved and (ii) since
$\Omega$ can be deduced algebraically from (\ref{sum}). As a
consequence, we retain the {\it closed} system
\begin{equation}\label{syst1}
\left\lbrace
\begin{array}{l}
\Omega_\Lambda'= 2(1+q)\Omega_\Lambda\\ \Omega_K'=2q\Omega_K
\end{array}
\right.
\end{equation}
with $q$ being a function of $\Omega_\Lambda$ and $\Omega_K$ only
and defined in (\ref{def_q}).

The system (\ref{syst1}) is {\it autonomous}~\cite{systdyn}, which
implies that there is a unique integral curve passing through a given
point, except where the tangent vector is not defined (fixed points). 
Note that at every point on the curve the system (\ref{syst1}) assigns
a unique tangent vector to the curve at that point.  It immediately
follows that two trajectories cannot cross; otherwise the tangent
vector at the crossing point would not be unique~\cite{systdyn}.

\subsection{Determination of the fixed points}\label{III2}

To study the system (\ref{syst1}) as a dynamical system, we first
need to determine the set of fixed points, i.e. the set of
solutions such that $\Omega_\Lambda'=0$ and $\Omega_K'=0$. These
solutions represent equilibrium positions which indeed can be
either stable or unstable. The fixed points are thus solutions of
\begin{equation}
 (1+q)\Omega_\Lambda=0,\qquad q\Omega_K=0.
\end{equation}
We obtain the three solutions
\begin{equation}
(\Omega_K,\Omega_\Lambda)\in\left\lbrace (0,0), (0,1),
(1,0)\right\rbrace.
\end{equation}
Each of these solutions represent a universe with different
physical characteristics:
\begin{enumerate}
\item\underline{$(\Omega_K,\Omega_\Lambda)=(0,0)$}: {\it the Einstein
de Sitter space} (EdS).

It is a universe with flat spatial sections, i.e. the three
dimensional hypersurfaces of constant time are Euclidean and it
has no cosmological constant. We deduce from (\ref{sum}) and
(\ref{def_q}) that
\begin{equation}
\Omega=1,\qquad q=\frac{3}{2}\gamma-1
\end{equation}
and integrating Eq. (\ref{fried1}) gives
\begin{equation}
a(t)=\left(\sqrt{\frac{\kappa C}{3}}t\right)^{\frac{2}{3\gamma}}
\end{equation}
for the solution vanishing at $t=0$. 

\item\underline{$(\Omega_K,\Omega_\Lambda)=(0,1)$}: {\it the de Sitter
space} (dS).

It is an empty space filled with a positive cosmological constant and
with flat spatial sections. We deduce from (\ref{sum}) and
(\ref{def_q}) that
\begin{equation}
\Omega=0,\qquad q=-1
\end{equation}
and integrating Eq. (\ref{fried1}) gives
\begin{equation}
a(t)= a_0\hbox{e}^{\sqrt{\frac{\Lambda}{3}}\,t}.
\end{equation}
This universe is accelerating in an eternal exponential expansion.

\item\underline{$(\Omega_K,\Omega_\Lambda)=(1,0)$}: {\it the Milne
universe} (M).

It is an empty space with no cosmological constant and with
hyperbolic spatial section ($K<0$).  We deduce from (\ref{sum})
and (\ref{def_q}) that
\begin{equation}
\Omega=0,\qquad q=0
\end{equation}
and integrating Eq. (\ref{fried1}) gives
\begin{equation}
a(t)= a_0 t
\end{equation}
for the solution vanishing at $t=0$ and in units where $K=a_0^2$.
\end{enumerate}

It is also interesting to study the properties of the three following
invariant lines which separate the phase space in disconnected regions:

\begin{enumerate}
\item\underline{$\Omega_K=0$}: The system (\ref{syst1}) reduces to
the equation of evolution for $\Omega_\Lambda$
\begin{equation}
\Omega_\Lambda'=3\gamma(1-\Omega_\Lambda)\Omega_\Lambda.
\end{equation}
Thus, if initially $\Omega_K=0$, we stay on this line during the whole
evolution and converge toward either $\Omega_\Lambda=1$ (i.e. the fixed
point dS) or toward $\Omega_\Lambda=0$ (i.e. the fixed point EdS). It
also follows that no integral flow lines of the system (\ref{syst1}) can
cross the line $\Omega_K=0$. It separates the universes with
$\Omega_K>0$ which are compact (i.e. having a finite spatial extension)
and the universes with $\Omega_K<0$ which are infinite (if one assumes
trivial topology~\cite{topo}). Crossing the line $\Omega_K=0$ would thus
imply a change of topology.  Note that if $\gamma=0$, the fluid behaves
like a cosmological constant and thus $\Omega_\Lambda'=0$ since
$\Omega_K$ remains zero.

\item\underline{$\Omega_\Lambda=0$}: The system (\ref{syst1}) reduces to
the equation of evolution for $\Omega_K$
\begin{equation}
\Omega_K'=(3\gamma-2)(1-\Omega_K)\Omega_K.
\end{equation}
As in the previous case, we stay on this line during the whole evolution
and converge toward either $\Omega_K=1$ (i.e. the fixed point M) or
toward $\Omega_\Lambda=0$ (i.e. the fixed point EdS).  It also follows
that no integral flow lines of the system (\ref{syst1}) can cross the
line $\Omega_\Lambda=0$.  Note that if $\gamma=2/3$, the fluid behaves
like a curvature term and thus $\Omega_K'=0$ since $\Omega_\Lambda$
remains zero.

\item\underline{$\Omega=0$}: It is a boundary of the phase space since
$\Omega$ is non negative. We now have $q=-\Omega_\Lambda$ and the system
(\ref{syst1}) reduces to 
\begin{equation}
\Omega_\Lambda'=2(1-\Omega_\Lambda)\Omega_\Lambda.
\end{equation}
The universe converges either toward (dS) or (M).
\end{enumerate}

\subsection{Stability analysis}\label{III3}

The second step is to determine whether these fixed points are
stable (i.e. attractors: A), unstable (i.e. repulsor: R) or saddle
(S) points (i.e. attractor in a direction and repulsor in
another). This property can be obtained by studying at the
evolution of a small deviation from the equilibrium
configuration. We thus decompose $\Omega_K$ and $\Omega_\Lambda$
as
\begin{eqnarray}
\Omega_K&\equiv&{\overline\Omega}_K+\omega_K\, ,\\
\Omega_\Lambda&\equiv&{\overline\Omega}_\Lambda+\omega_\Lambda\, ,
\end{eqnarray}
where $({\overline\Omega}_\Lambda,{\overline\Omega}_K)$ represents
the coordinates of one of the fixed points determined in the
previous section and where $(\omega_\Lambda,\omega_K)$ is a small
deviation around this point.

Writing the system of evolution (\ref{syst1}) as
\begin{equation}
\left(\begin{array}{c} \Omega_K
\\ \Omega_\Lambda\end{array}\right)'=
\left(\begin{array}{c} F_K(\Omega_\Lambda,\Omega_K)
\\ F_\Lambda(\Omega_\Lambda,\Omega_K)\end{array}\right),
\end{equation}
where $F_K$ and $F_\Lambda$ are two functions
determined from (\ref{syst1}), it can be expanded to linear order
around $({\overline\Omega}_K,{\overline\Omega}_\Lambda)$ (for
which $F_K$ and $F_\Lambda$ vanish) to give the equation of
evolution of $(\omega_\Lambda,\omega_K)$
\begin{equation}
\left(\begin{array}{c} \omega_K
\\ \omega_\Lambda\end{array}\right)'=
\left(\begin{array}{cc} \frac{\partial F_K}{\partial\Omega_K} &
\frac{\partial F_K}{\partial\Omega_\Lambda} \\ \frac{\partial
F_\Lambda}{\partial\Omega_K} & \frac{\partial
F_\Lambda}{\partial\Omega_\Lambda}
\end{array}\right)_{({\overline\Omega}_\Lambda,{\overline\Omega}_K)}
\left(\begin{array}{c} \omega_K
\\ \omega_\Lambda\end{array}\right)
\equiv{\cal P}_{({\overline\Omega}_\Lambda,{\overline\Omega}_K)}
\left(\begin{array}{c} \omega_K
\\ \omega_\Lambda\end{array}\right).
\end{equation}
The stability of a given fixed point depends on the sign of the two
eigenvalues ($\lambda_{1,2}$) of the matrix ${\cal
P}_{({\overline\Omega}_\Lambda,{\overline\Omega}_K)}$. If both
eigenvalues are positive (resp. negative) then the fixed point is a
repulsor (resp. an attractor) since $(\omega_K,\omega_\Lambda)$ will
respectively goes to infinity (resp. zero). In the case where the two
eigenvalues have different signs, the fixed point is an attractor
along the direction of the eigenvector associated with the negative
eigenvalue and a repulsor along the direction of the eigenvector
associated with the positive eigenvalue. We also introduce ${\bf
u}_{\lambda_{1,2}}$ the eigenvectors associated to the two eigenvalues
which give the (eigen)--directions of attraction or repulsion.

We have to perform this stability analysis for each of the three
fixed points (reminding that $\gamma\not=0,2/3)$:
\begin{enumerate}
\item\underline{EdS fixed point}: In that case, the matrix ${\cal
P}$ is given by
\begin{equation}
{\cal P}_{\rm EdS}=\left(
\begin{array}{cc}
3\gamma-2 & 0 \\ 0 & 3\gamma
\end{array}\right),
\end{equation}
the eigenvalues of which are trivially given by $3\gamma-2$ and
$2\gamma$. We thus conclude that if $\gamma\in]-\infty,0[$ then
EdS is an attractor, that it is a saddle point when
$\gamma\in]0,2/3[$ and a repulsor when $\gamma\in]2/3,+\infty[$.
The matrix ${\cal P}_{\rm EdS}$ being diagonal the two eigenvectors
are trivially given by
\begin{equation}
{\bf u}_{(3\gamma)}=(0,1),\qquad {\bf u}_{(3\gamma-2)}=(1,0)
\end{equation}
corresponding respectively to two invariant boundaries
$\Omega_\Lambda=0$ and $\Omega_K=0$.
\item\underline{dS fixed point}: The matrix ${\cal P}$ is now given by
\begin{equation}
{\cal P}_{\rm dS}=\left(
\begin{array}{cc}
-2 & 0 \\ 2-3\gamma & -3\gamma
\end{array}\right).
\end{equation}
The eigenvalues of ${\cal P}_{\rm dS}$ are $-2$ and $-3\gamma$. It
follows that the fixed point dS is never a repulsor. If
$\gamma\in]-\infty,0[$ then dS is a saddle point and, when
$\gamma\in]0,+\infty[$, it is an attractor.  The two eigenvectors are
now given by
\begin{equation}
{\bf u}_{(-3\gamma)}=(0,1),\qquad {\bf u}_{(-2)}=(1,-1)
\end{equation}
corresponding respectively to the two boundaries $\Omega=0$ and
$\Omega_\Lambda=0$.

\item\underline{M fixed point}:  The matrix ${\cal P}$ is now given by
\begin{equation}
{\cal P}_{\rm M}=\left(
\begin{array}{cc}
2-3\gamma & -3\gamma \\ 0 & 2
\end{array}\right).
\end{equation}
The eigenvalues of ${\cal P}_{\rm M}$ are $2$ and $2-3\gamma$. It
follows that M is never an attractor since one of his eigenvalues is
always positive. If $\gamma\in]-\infty,2/3[$ then M is a repulsor point
and, when $\gamma\in]2/3,+\infty[$, it is a saddle point. The two
eigenvectors are now given by
\begin{equation}
{\bf u}_{(2-3\gamma)}=(1,0),\qquad {\bf u}_{(2)}=(1,-1)
\end{equation}
corresponding respectively to the two boundaries $\Omega_K=0$ and
$\Omega=0$.
\end{enumerate}

Before we sum up all theses results, let us concentrate about the
cases where $\gamma=0$ or $\gamma=2/3$ in which the matter behaves
respectively either as a cosmological constant or as a curvature
term. As a consequence $\Omega$ can be absorbed in a redefinition
of either $\Omega_\Lambda$ or $\Omega_K$ and we can set $\Omega=0$
from which it follows that (\ref{sum}) implies
$\Omega_\Lambda+\Omega_K=1$. In both cases, we deduce from
(\ref{def_q}) that $q=\Omega_K-1=-\Omega_\Lambda$ so that
\begin{equation}
\Omega_K'=2(\Omega_K-1)\Omega_K,\qquad
\Omega_\Lambda'=2(1-\Omega_\Lambda)\Omega_\Lambda
\end{equation}
which are not independent equations due to the constraint
$\Omega_\Lambda+\Omega_K=1$. Thus, for $\gamma=0$ or $\gamma=2/3$,
the two fixed points are either (M) or (dS) which are respectively
a repulsor and an attractor.

\begin{figure}
\begin{center}
\epsfig{figure=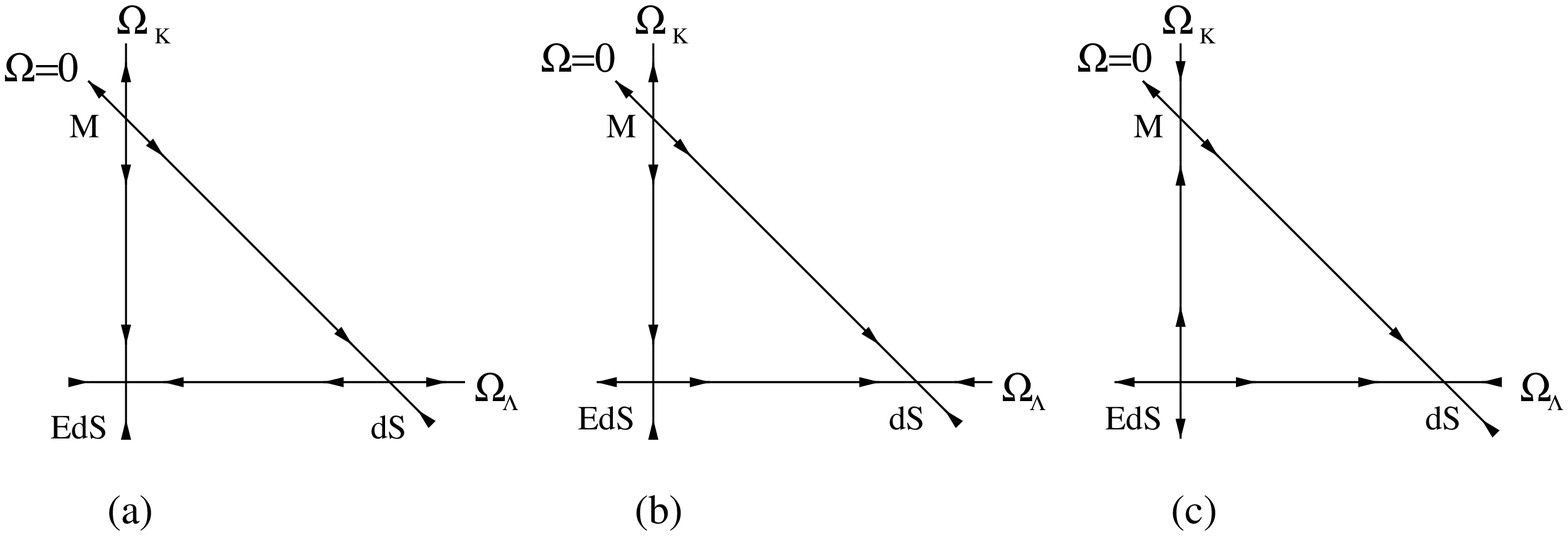, width=14cm}
\end{center}
\caption{The fixed points and their stability depending of the value
of the index $\gamma$: (a) $\gamma<0$, (b) $0\leq\gamma<2/3$
and (c) $\gamma\geq2/3$.}\label{fig0}
\end{figure}

As a conclusion of this study, we sum up the properties of the three
spacetimes as a function of the polytropic index of the cosmic fluid in
table~\ref{tab1} and in figure~\ref{fig0}, we depict the fixed points,
their directions of stability and instability as well as the invariant
boundary in the plane ($\Omega_\Lambda,\Omega_K$). Indeed, the attractor
solution can be guessed directly from Eq.~(\ref{fried1}) and the
behavior (\ref{Cgamma}) of the density with the scale factor since if
$\gamma<0$ the matter energy density scales as $a^{-3\gamma}$ and comes
to dominate over the cosmological constant (scaling as $a^0$) and the
curvature (scaling as $a^{-2}$). On the other hand the cosmological
constant always finishes by dominating if $\gamma>0$. The curvature can
never dominates in the long run since it will be caught up by either the
matter or the cosmological constant.

\begin{table}
\caption{Stability properties of the three fixed points (EdS, dS and M)
         as a function of the polytropic index $\gamma$. 
         (A: attractor, R: repulsor and S: saddle point)}
\label{tab1}
\begin{indented}
\item[]
\begin{tabular}{lccccc}
\br
$\gamma$ & $]-\infty,0[$ & 0 & $]0,2/3[$ & $2/3$ &
$]2/3,+\infty[$\\
\mr
EdS & A & N.A. & S & N.A. &R\\
dS& S & A & A & A & A\\
M & S & R & R & R& S\\
\br
\end{tabular}
\end{indented}
\end{table}

\section{Numerical examples}\label{IV}

The full phase space picture can be obtained only through a
numerical integration of the system (\ref{syst1}) by using an
implicit fourth order Runge--Kutta method~\cite{numeric}.

\begin{figure}
\begin{center}
\epsfig{figure=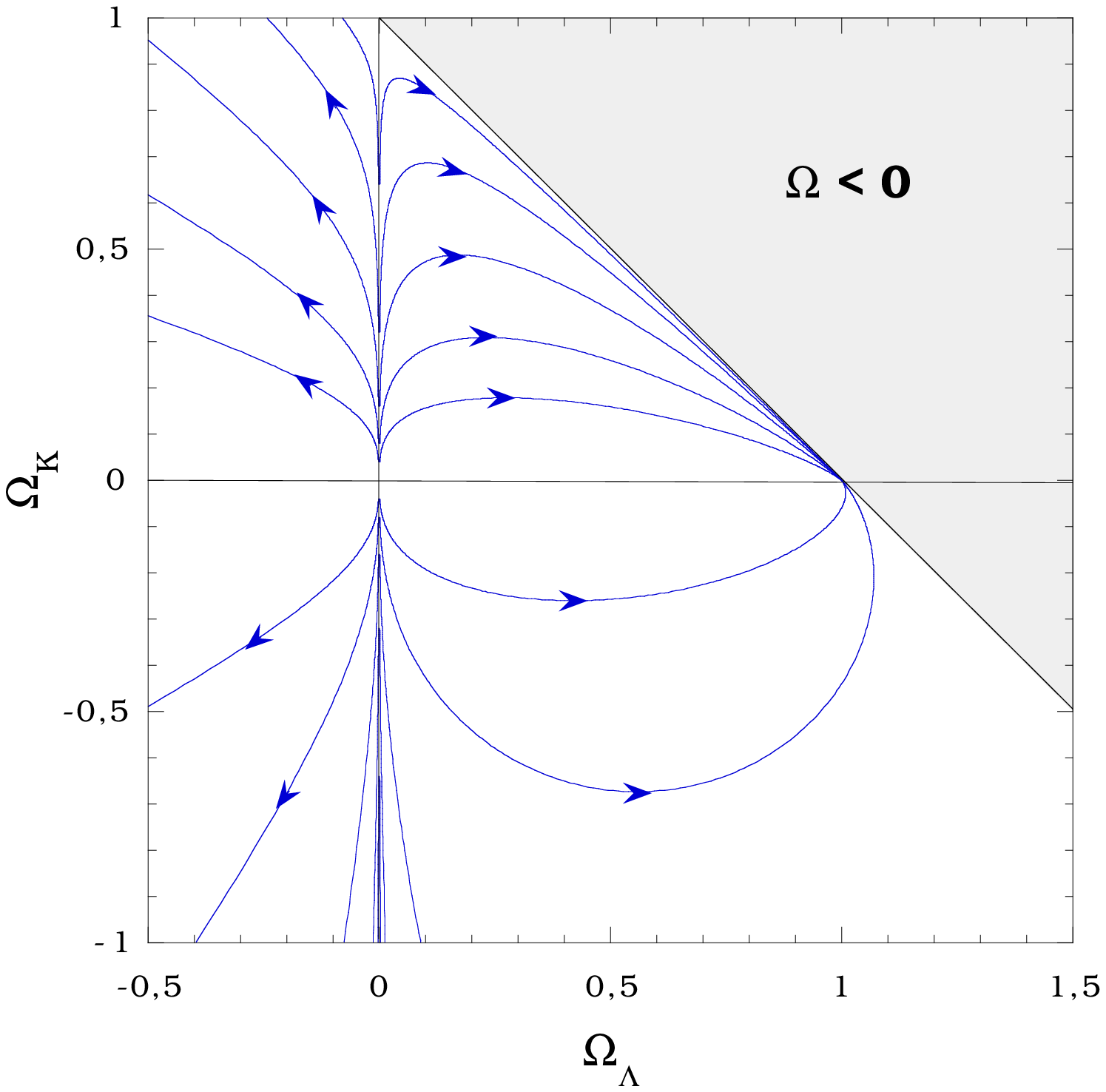, width=7cm}
\epsfig{figure=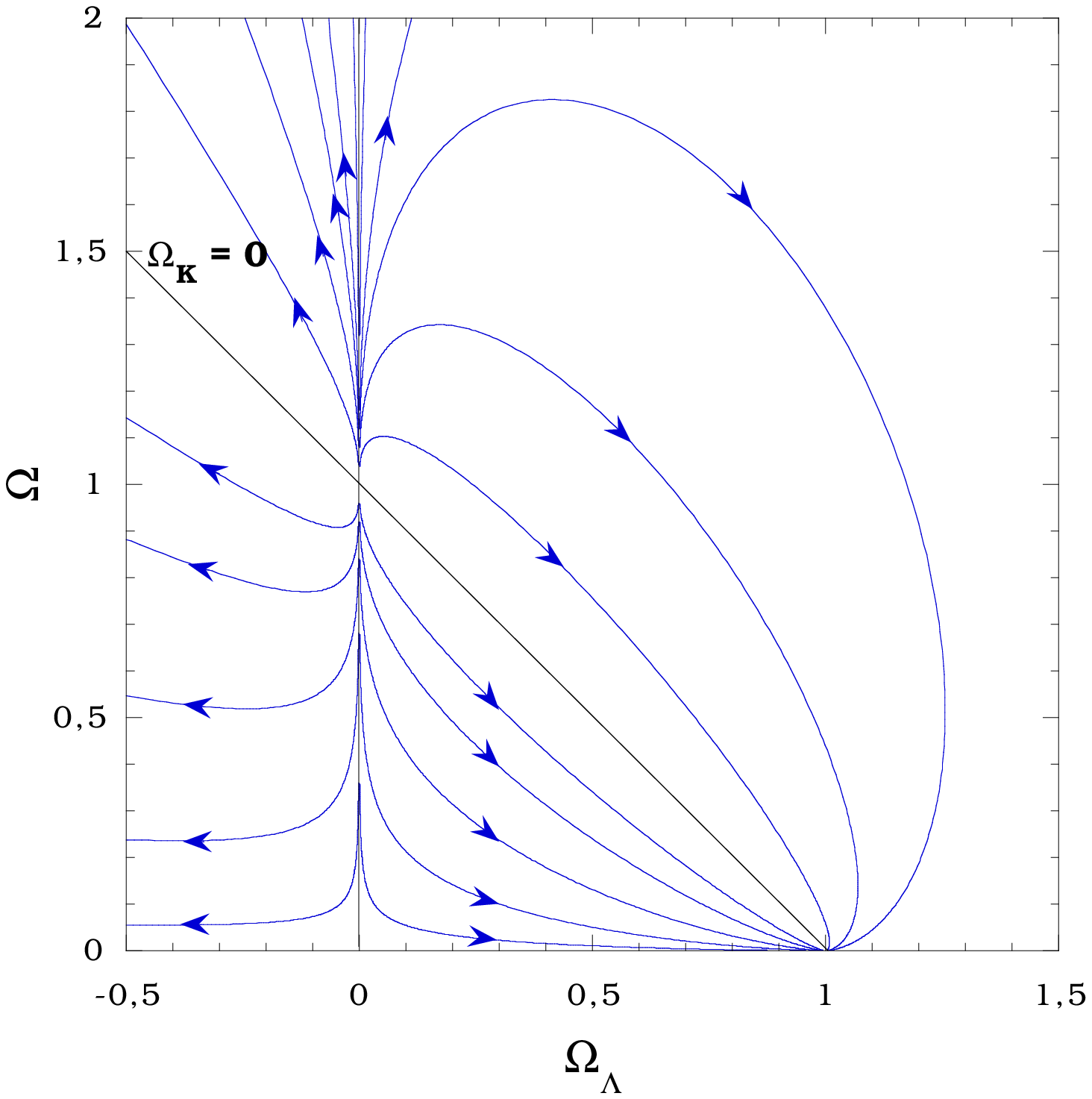, width=7cm}
\end{center}
\caption{Phase space of the system (\ref{syst1}) in the plane
$(\Omega_K,\Omega_\Lambda)$ [left] and $(\Omega,\Omega_\Lambda)$
[right]. We have represented the three fixed points and the lines
$\Omega_K=1$ and $\Omega_\Lambda=1$ and we have considered the
value $\gamma=1$ (i.e. pressureless fluid). }\label{fig1}
\end{figure}

\begin{figure}
\begin{center}
\epsfig{figure=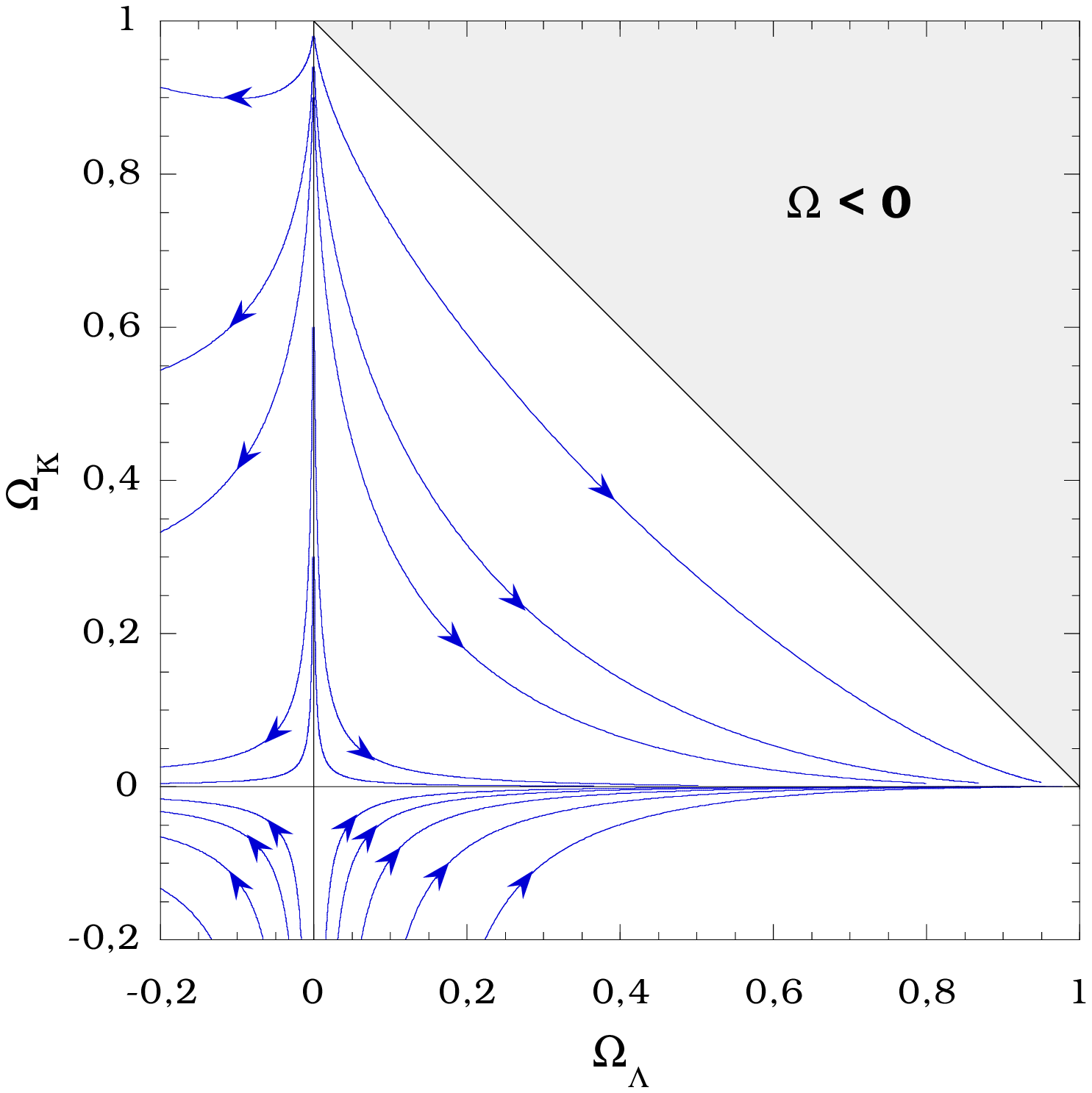, width=7cm}
\epsfig{figure=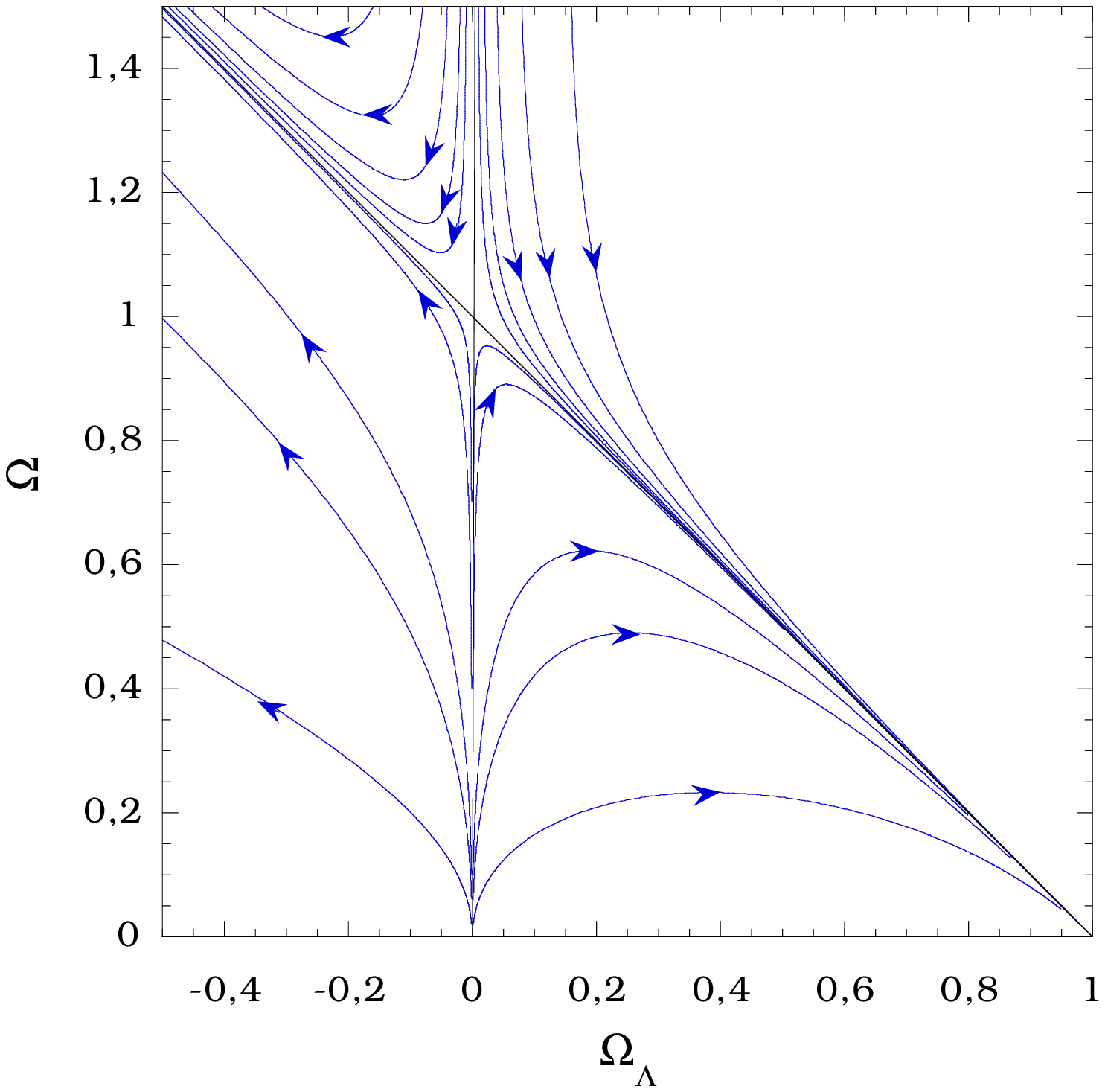, width=7cm}
\end{center}
\caption{Phase space of the system (\ref{syst1}) in the plane
$(\Omega_K,\Omega_\Lambda)$ [left] and $(\Omega,\Omega_\Lambda)$
[right] for $\gamma=1/3$. }\label{fig2}
\end{figure}

\begin{figure}
\begin{center}
\epsfig{figure=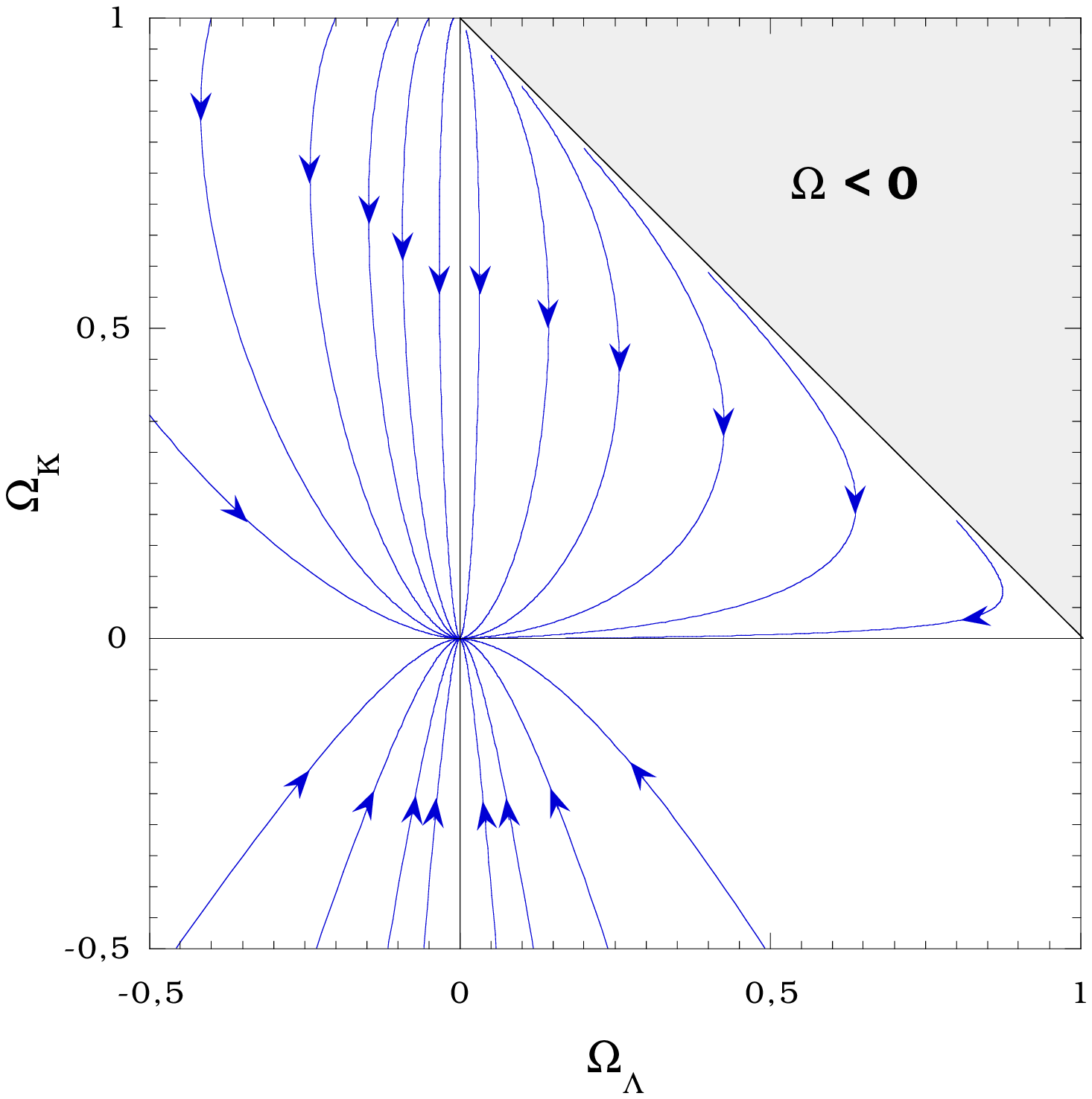, width=7cm}
\epsfig{figure=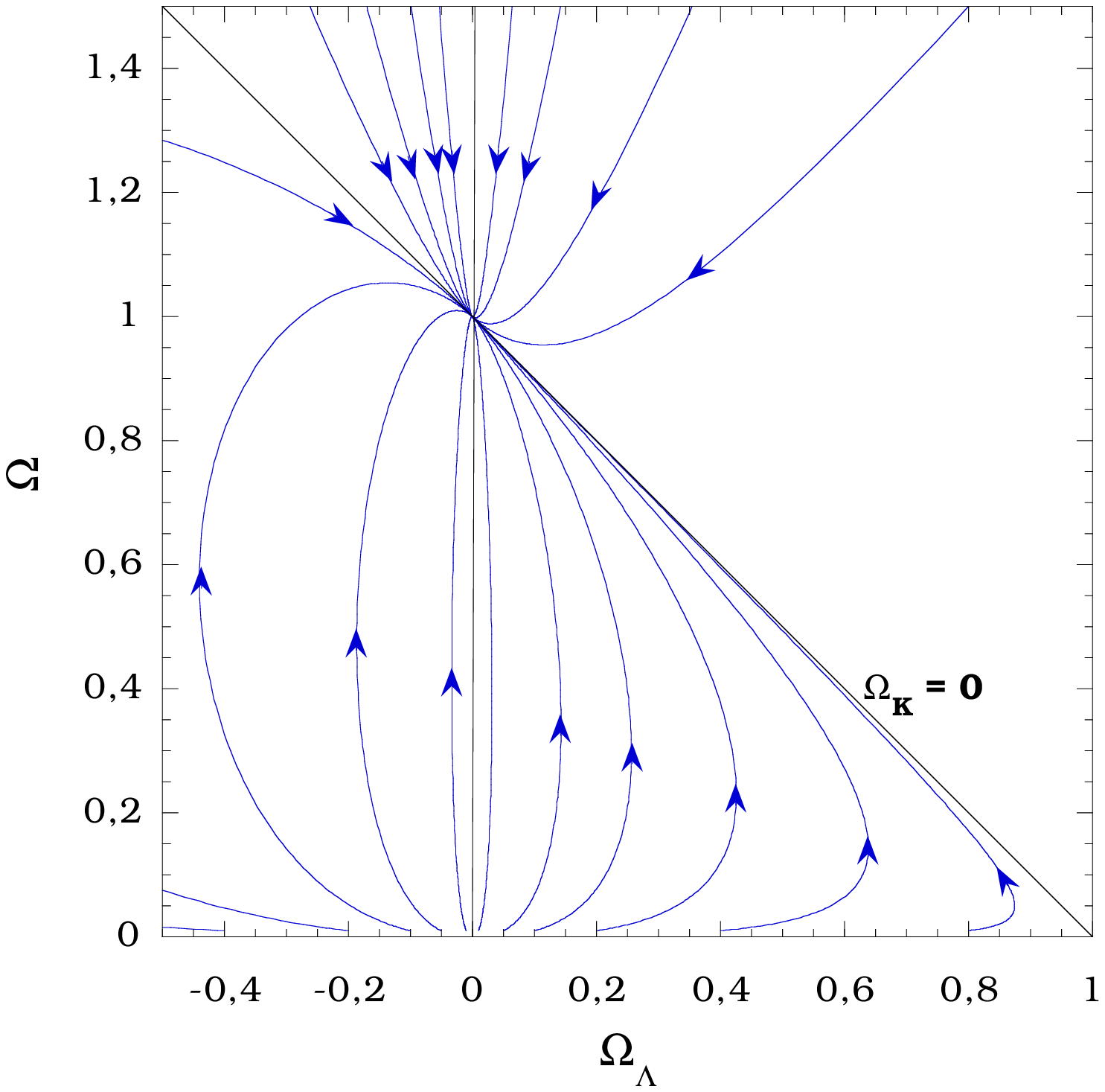, width=7cm}
\end{center}
\caption{Phase space of the system (\ref{syst1}) in the plane
$(\Omega_K,\Omega_\Lambda)$ [left] and $(\Omega,\Omega_\Lambda)$
[right] for $\gamma=-1$. }\label{fig3}
\end{figure}

Ordinary matter such as a pressureless fluid or a radiation fluid has
$\gamma>1$ and we first consider this case on figure~\ref{fig1} where we
depict the phase space both in the $(\Omega_K,\Omega_\Lambda)$ where the
analytic study of the fixed points was performed but also in the plane
$(\Omega,\Omega_\Lambda)$ for complementarity.  On figure~\ref{fig2}, we
consider the case where $0<\gamma<2/3$ which can corresponds to a scalar
field slowly rolling down its potential or a tangle of domain strings
(for which $\gamma=1/3$) and we finish by the more theoretical case where
$\gamma<0$ on figure~\ref{fig3} for which we know no simple physical 
example (see however \cite{prdjpu}).


\section{Discussion and conclusions}\label{V}

To discuss the naturalness of the initial conditions leading to
our observed universe, we have to add the actual observational
measures in the plane ($\Omega,\Omega_\Lambda$) and trace them
back to estimate the domain in which our universe has started.
This required (i) to know what are the constraints on the
cosmological constant and the curvature of the universe and (ii)
determine the age of the universe, i.e. the time during which we
must integrate back.

It is not the purpose of this article to detail the observational
methods used in cosmology and a description can be found en
e.g.~\cite{peebles93}; we now just sum up what is thought to be the
current status of these observations.  The current observational data
such as the cosmic microwave background measurements~\cite{cmb}, the
Type Ia supernovae data~\cite{sn1}, large scale velocity
fields~\cite{velo}, gravitational lensing~\cite{gravlens} and the
measure of the mass to light ratio~\cite{cluster} tend to show that
\begin{equation}
\Omega_0\sim0.3,\quad\Omega_{\Lambda_0}\sim0.7.
\end{equation}
We refer the reader to the review by Bahcall {\em et al.}~\cite{bahcall}
for a combined study of these data and a description of all the
observation methods. Let us just keep in mind that we are
close to the line $\Omega_K=0$ and let us consider the safe area of parameter
such that
\begin{equation}
{\cal D}_0:\quad
\left\lbrace\Omega_0\in[0.1,0.5],\quad\Omega_{\Lambda_0}\in[0.5,0.9]
\right\rbrace 
\end{equation}
and let us determinate the initial conditions allowed by these
observations.

\begin{figure}
\begin{center}
\epsfig{figure=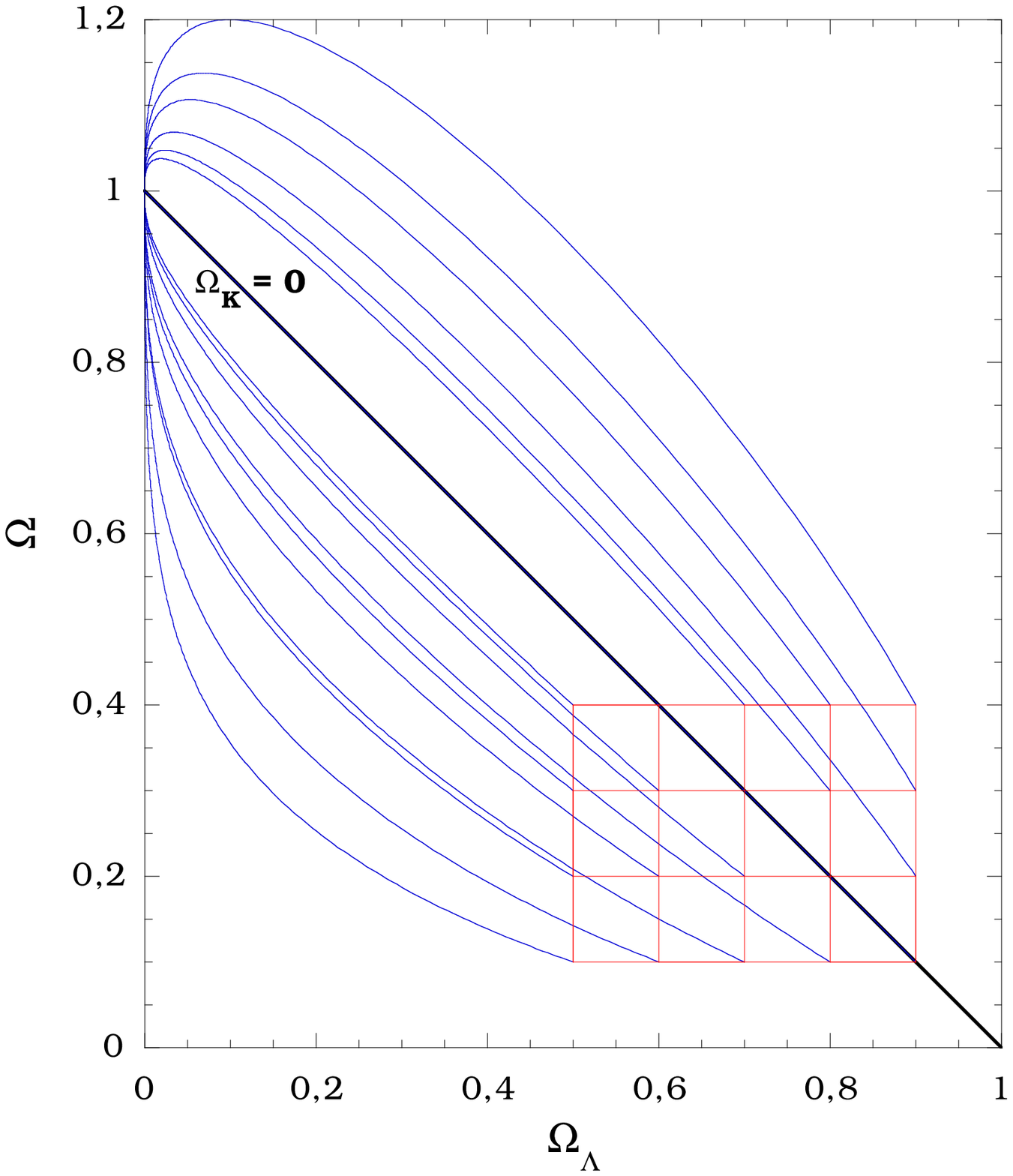, width=7cm}
\epsfig{figure=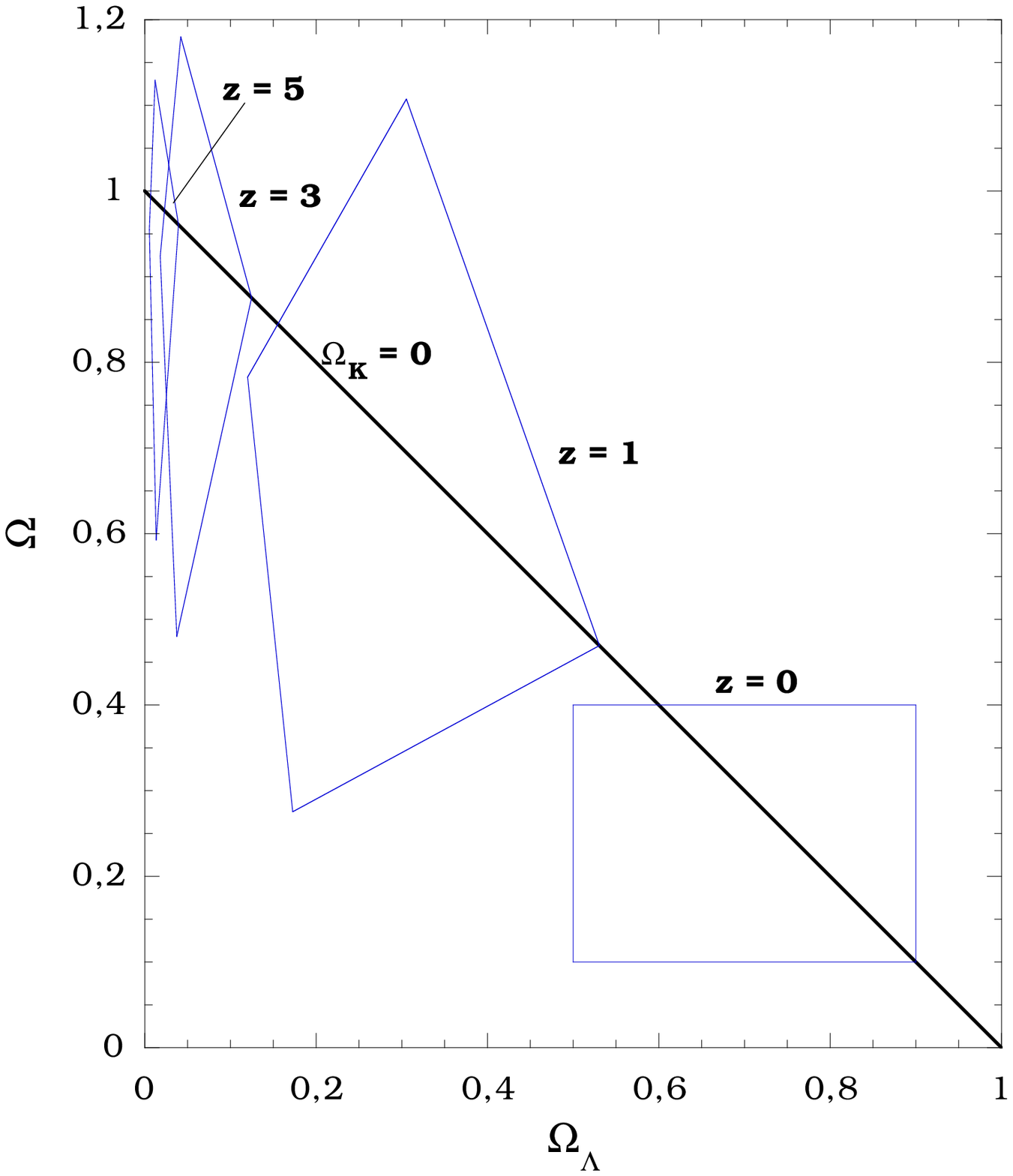, width=7cm}
\end{center}
\caption{The domain ${\cal D}_0$ of current observational values for our
universe and its inverse image by the system (\ref{syst1}) when one is
assuming that the universe has been dominated by a pressureless fluid
during its whole evolution.}\label{fig5}
\end{figure}

For that purpose, we need to integrate the system (\ref{syst1}) back in
time during a time equal to the age of the universe. Today, the matter
content of the universe is dominated by a pressureless fluid, the energy
density of which is obtained once $\Omega_0$ has been chosen and is
\begin{equation}
\rho_{\rm mat}=\frac{3H_0^2}{\kappa}\Omega_0\left(\frac{a}{a_0}\right)^{-3} =
1.80\times 10^{-29}\Omega_0\,h^2\, \left(\frac{a}{a_0}\right)^{-3}
\,\hbox{g.cm}^{-3}
\end{equation}
where $H_0=100\,h$ km/s/Mpc is the Hubble constant today. The energy
density of the radiation is obtained by computing the energy contained
in the cosmic microwave background which is the dominant contribution to
the radiation in the universe. Since it is a black body with temperature
$\Theta_0=2.726\,$K, we deduce, from the Stephan-Boltzmann law, that
\begin{equation}
\rho_{\rm rad}=4.47(1+f_\nu)\times10^{-34}\,\left(\frac{a}{a_0}\right)^{-4}
\hbox{g.cm}^{-3}
\end{equation}
where $f_\nu=0.68$ is a factor to take into account the contribution of
three families of neutrinos~\cite{peebles93}. The radiation was thus
dominating over the matter for scale factors smaller than $a_{\rm eq}$
at which $\rho_{\rm mat}=\rho_{\rm rad}$ and thus given by
\begin{equation}
\frac{a_{\rm eq}}{a_0}\simeq\frac{4.5\times10^{-5}}{\Omega_0\,h^2}.
\end{equation}
We can integrate back until the Planck era for which $a_{\rm Pl}/a_0\sim
10^{-30}$ and can thus approximate $\gamma$ by
\begin{equation}
\gamma=\left\lbrace
\begin{array}{ll}
1
&a_{\rm eq}\leq a\leq a_0\\
4/3
&a_{\rm Pl}\leq a\leq a_{\rm eq}
\end{array}\right.
\end{equation}
which is a good approximation for $\gamma=1+1/3(1+a/a_{\rm eq})$.  In
figure~\ref{fig5}, we depict the domain ${\cal D}_0$ of current
observational values and its inverse image by the system (\ref{syst1})
at the beginning of the matter era and at the end Planck era.

To illustrate this fine tuning problem analytically, let us just
consider the simplest case where $\Omega_\Lambda=0$ for which the
evolution of $\Omega$ is simply given by
\begin{equation}
\Omega'=(3\gamma-2)\Omega(\Omega-1)
\end{equation}
the solution of which is
\begin{equation}
\Omega=\frac{1}{1+\frac{\Omega_{K_0}}{\Omega_0}\left(\frac{a}{a_0}
\right)^{3\gamma-2}} 
\end{equation}
and thus,
\begin{equation}
\Omega=\left\lbrace
\begin{array}{ll}
\frac{1}{1+\frac{\Omega_{K_0}}{\Omega_0}\frac{a}{a_0}}
&a_{\rm eq}\leq a\leq a_0\\
\frac{1}{1+\frac{\Omega_{K_0}}{\Omega_0}\frac{a_0}{a_{\rm eq}}
\left(\frac{a}{a_0}\right)^2}
&a_{\rm Pl}\leq a\leq a_{\rm eq}
\end{array}\right.
..
\end{equation}
From the observational data we get that $\Omega_{K_0}\sim{\cal
O}(10^{-1})$ and thus  $\Omega_{0}\sim{\cal O}(1)$ from which we
deduce that
\begin{equation}
\left.\Omega_{K}\right|_{a=a_{\rm eq}}\sim {\cal O}(10^{-4}),\qquad
\left.\Omega_{K}\right|_{a=a_{\rm Pl}}\sim {\cal O}(10^{-58}).
\end{equation}
This illustrate that a almost flat universe requires a fine tuning of
the curvature, which is a consequence that $\Omega_K=0$ is a repulsor in
both the radiation and the matter era.

A solution to solve this fine tuning problem would be to add a phase
prior to the radiation era in which $\gamma\leq 0$ so that  $\Omega_K=0$
becomes an attractor and to tune the duration of this era such as
to have the correct initial conditions.  Then, during the standard
evolution EdS becomes repulsor and we evolve toward dS staying close
to the line $\Omega_K = 0$ hence explaining current observations. 
Inflation is a realisation of such a scenario.  What inflation really
does is to change the stability property of the fixed points and
invariant boundaries of the system (\ref{syst1}).  Hence during this
period where a fluid with negative pressure is dominating we are
attracted close to the line $\Omega_K = 0$ and the closer the longer this
phase lasts.  We then switch to a phase with normal matter and
start to drift away due to repulsive property of EdS. Nevertheless,
inflation does more than just explaining where we stand in this phase
space, it also gives an explanation for the observed structures
(galaxies, clusters...)  of our universe, but this is beyond the scope
of this article.

\ack{We thank Lucille Martin and Jean-Pierre Luminet for discussions and
Brandon Carter for clarification concerning the invariance of Newtonian
mechanics. JPU dedicates this work to Yakov.}\\

\end{document}